\documentclass[12pt]{article}
\usepackage{amsmath}
\usepackage{amsfonts}
\usepackage{amsthm}
\usepackage{graphicx}
\usepackage{mathrsfs}
\catcode`@=11   
\newcommand{\eq}{\begin{equation}}
\newcommand{\en}{\end{equation}}
\newcommand{\eqn}{\begin{eqnarray}}
\newcommand{\enn}{\end{eqnarray}}

\newcommand{\beq}{\begin{equation}}
\newcommand{\eeq}{\end{equation}}

\def\CP#1{\relax\ifmmode\IP^{#1}\else\IP$^{#1}$\fi}

\def\IP{\relax\leavevmode{\rm I\kern-.18em P}}

\def\Ione{\relax\leavevmode{\rm 1\kern-3pt l}}

\def\sgn{\mathop{\operator@font sign}\nolimits}

\let\7=\widetilde

\def\IR{\relax\leavevmode{\rm I\kern-.18em R}}
\def\ZZ{\relax\leavevmode
                   \ifmmode\mathchoice
                   {\hbox{\sf Z\kern-.4em Z}}
                   {\hbox{\sf Z\kern-.4em Z}}
                   {\lower.9pt\hbox{\scriptsize\sf Z\kern-.36em Z}}
                   {\lower1.2pt\hbox{\tiny\sf Z\kern-.36em Z}}
                    \else{\sf Z\kern-.4em Z}\fi}
\def\RR{\relax\leavevmode
                   \ifmmode\mathchoice
                   {\hbox{\sf R\kern-.4em R}}
                   {\hbox{\sf R\kern-.4em R}}
                   {\lower.9pt\hbox{\scriptsize\sf R\kern-.36em R}}
                   {\lower1.2pt\hbox{\tiny\sf R\kern-.36em R}}
                    \else{\sf R\kern-.4em R}\fi}

\def\resetby#1#2{\@addtoreset{#2}{#1}}
\def\seceq{\@addtoreset{equation}{section}
             \def\theequation{\thesection.\arabic{equation}}} 

\def\Label#1{\label{#1}%
         \smash{\hbox to0pt{\raise1ex\hbox{\tiny[#1]}\hss}}}
\def\noLabels{\let\Label=\label}

\reversemarginpar
\noLabels 
\catcode`@=12                   
\usepackage{amssymb}

\begin{document}

\begin{titlepage}
\begin{flushright}
\end{flushright}

\begin{center}

{\Large\bf{
              Nonassociativity, Malcev Algebras and
              String Theory}}\\[10mm]
{\bf M. G\"{u}naydin\footnote{e-mail: murat@phys.psu.edu} } \\[1mm]
            Institute for Gravitation and the Cosmos and Physics Department, \\
             Penn State University, 
             University Park, PA 16802\\[5mm]
{\bf D. Minic\footnote{e-mail: dminic@vt.edu} } \\[1mm]
             Department of Physics,  Virginia Tech, 
             Blacksburg, VA 24061\\
\vspace{0.5cm}
{\it \bf Dedicated to Bruno Zumino on the occasion of his 90th birthday}
\vspace{0.5cm}\\
{\bf Abstract}\\[4mm]
\parbox{4.8in}{
Nonassociative structures have appeared in the study of D-branes in curved backgrounds. In recent work,  string theory backgrounds involving three-form fluxes, where such structures show up, have been studied in more detail. We point out
that under certain assumptions these nonassociative structures coincide
with nonassociative Malcev algebras which had appeared  in
the quantum mechanics of systems with non-vanishing three-cocycles, such as a  point particle moving in the field of   a magnetic charge.  We generalize the corresponding Malcev algebras  to include electric as well as magnetic charges. These structures find
their classical counterpart in the theory of Poisson-Malcev algebras and their generalizations. We also study their connection to  Stueckelberg's generalized Poisson brackets that do not obey the Jacobi identity and point out that nonassociative string theory with a fundamental length corresponds to a realization  of his goal to find  a non-linear extension of quantum mechanics with a fundamental length. 
Similar nonassociative structures are also known to appear in the cubic formulation of closed
string field theory in terms of open string fields, leading us to conjecture a natural string-field theoretic
generalization
of the AdS/CFT-like (holographic) duality.
}


\end{center}

\end{titlepage}

\section{Introduction}

Noncommutative and associative algebraic structures are hallmarks of quantum
physics. In the Hilbert space formulation of quantum mechanics the linear
operators, such as  Hermitian operators corresponding to observables or
 generators of symmetries do not, in general, commute. Their actions on
the  states in the Hilbert space are however associative. 

In the early period
of quantum mechanics Pascual Jordan introduced the Jordan formulation of
quantum mechanics with the hope of generalizing the underlying algebraic
framework of quantum mechanics \cite{jordan}. Since the commutator of two Hermitian operators is not Hermitian Jordan suggested working with the symmetric product that preserves Hermiticity. The linear operators acting
in a Hilbert space form a commutative but nonassociative Jordan  algebra under the  symmetric product defined
as one  half the anti-commutator.  Jordan algebras that can
be realized as  such are called special. Jordan's hope of finding a large
class of Jordan algebras that are not special was dashed when he, von Neumann
and Wigner classified all finite dimensional Jordan algebras \cite{jordan}.
They showed that with but one exception all finite dimensional Jordan algebras
are special. The only Jordan algebra that has no realization in terms of
associative operators acting on a vector space or Hilbert space with the Jordan product being
one-half the anti-commutator is the algebra of $3\times 3$ Hermitian matrices
over the octonions which was called the  exceptional Jordan algebra and is denoted as $J_3^{\mathbb{O}}$. 

It was
shown in \cite{gpr} that one can formulate quantum mechanics over the exceptional
Jordan algebra, $J_3^{\mathbb{O}}$, satisfying all the axioms of quantum mechanics as  formulated
by von Neumann. This  work answered in affirmative the question posed by
von Neumann four decades earlier as to whether there exists a quantum mechanics
whose projective geometry is non-Desarguian.  The quantum mechanics defined
by the exceptional Jordan algebra is referred to as  octonionic quantum mechanics
and it does not admit a Hilbert space formulation.

In the eighties Zelmanov proved that there do not exist
any new non-special infinite dimensional Jordan algebras \cite{Zelmanov}.
This rules out the existence of an infinite dimensional quantum mechanics
that has no Hilbert space formulation. Hence the octonionic quantum mechanics
stands alone as the only quantum mechanics that has no Hilbert space formulation.
The fact that its geometry  is non-Desarguian implies that it can not be
embedded in a higher dimensional projective geometry \cite{mg_jhw}. 

The Jordan's formulation
of quantum mechanics in terms of the symmetric Jordan product was generalized
to a formulation in terms of quadratic Jordan algebras which extends to the
octonionic quantum mechanics as well as to quantum mechanics over finite
fields \cite{mg}. 

Remarkably, there exists a unique supergravity theory defined by the exceptional Jordan algebra \cite{GST} that is referred as the exceptional supergravity. In five space-time dimensions it describes the coupling of 14 Abelian vector multiplets to $N=2$ supergravity and is the largest of four magical  Maxwell-Einstein supergravity theories (MESGT). The magical supergravity theories are the only unified $5D$ MESGTs whose scalar manifolds are symmetric  spaces \cite{GST}. The exceptional  supergravity has the groups of the $E$-series as its global symmetry group in five, four and  three dimensions just like the maximal supergravity. However,  the  real forms of the noncompact global symmetry groups of the exceptional supergravity with eight supercharges are different from the real forms of the global symmetry groups of maximal supergravity with 32 supercharges in the respective dimensions.   Global symmetry groups of the exceptional supergravity in these dimensions are defined by the underlying exceptional Jordan algebra $J_3^{\mathbb{O}}$ over the real octonions $\mathbb{O}$. On the other hand the global symmetries of the maximal supergravity in the respective dimensions are defined by the split exceptional Jordan algebra $J_3^{\mathbb{O_S}}$ over split octonions $\mathbb{O}_S$\footnote{See the review lectures  \cite{Gunaydin:2009pk} and the references therein.}.

Among the issues investigated by the founders of quantum mechanics
was the question whether quantum mechanics requires the use of complex number
field in its formulation.
In a remarkable paper on the question of whether one can formulate  quantum
theory over the field of real numbers \cite{ecgs} 
Stueckelberg showed that one can generalize  Poisson brackets in classical
mechanics such that they no longer satisfy the Jacobi identity while still
preserving 
the Liouville theorem and  the fundamental $H$-theorem of Boltzmann.
This result led him to ask the question whether there exists a nonlinear
generalization of quantum mechanics that involves a fundamental length, corresponding to his generalized Poisson brackets.

 Nambu \cite{triplenambu} pointed out that there exists a natural generalization of the canonical Poisson bracket that involves a triple product, which is surprisingly non-trivial
to quantize and  seems to be relevant for the covariant formulation of M-theory \cite{nambu}. Nambu's work was partly inspired by the work of G\"unaydin and G\"ursey on the implications of extending the underlying division algebra of quantum mechanics to octonions and the connection between color degrees of freedom of quarks and octonions \cite{muratfeza}.\footnote{
Note that the role of nonassociativity is not fundamental in the context of the Nambu bracket and that quantization can be carried out with
standard Hilbert space methods \cite{curtright}.}

It was also argued that nonassociative structures appear naturally in closed string field theory \cite{horstrom, strom, strom1}.
Other physicists have persisted in being fascinated
by the possibility of nonassociative structures (see the monographs
\cite{okubo}, \cite{gursey} for a more complete set of references on the
subject).

In spite of all these efforts, it is fair to say that we still do not know what precise role intrinsic nonassociativity will play in a deeper understanding
of fundamental physics, in general, and in the foundations of M/superstring theory, in particular.
In this note, we point out the relevance of certain specific non-associative
structures, such as Malcev algebras,  to  string  theory and to quantum mechanical extensions of Stueckelberg's generalized Poisson brackets.

It is well known that the constant $B$-field background in string theory leads to a non-commutative but
associative star product
\cite{kont}. As a consequence, the Seiberg-Witten limit of open string field theory then
leads to a noncommutative gauge theory \cite{connes, seibwitten, schom, chu}.

Following these classic results, it has been pointed out that the presence of a non-zero $H=dB$ background can give rise to
nonassociative star-products \cite{corn, ho}. More recently conformal field theories
involving the three-form backgrounds have been studied in \cite{lust}. Thus nonassociative structure
seem to find natural appearance in closed string theory after all. This is perhaps not surprising
given the pioneering papers \cite{horstrom, strom, strom1}.

One of the main goals of  this paper is to point out that in the case when $H$ is constant, the  physical
situation is well known from the study of non-vanishing three-cocycles in quantum
mechanics \cite{threecyc}. The canonical example is that of a non-relativistic electron
moving in the  field of a constant magnetic charge distribution \cite{lipkin}, which is relevant for closed string theory in the constant $H$ background \cite{lust}. As was pointed out in
\cite{guzu} the relevant nonassociative structure in the case of constant magnetic charge is the so called Malcev
algebra \cite{malcev, sagle, loos, allison}. Thus the nonassociative algebraic
structures which appear in closed string theory in the constant $H$ background  \cite{lust} are
Malcev algebras. However, these structures need to be generalized in consistent string
backgrounds with constant $H$ fields, due to the presence of a non-trivial gravitational background.
We also point out that these quantum structures have a natural classical limit, in which they are 
precisely related to the prescient work of Stueckelberg, reviewed in section 2.

Motivated by the discussion of strings in non-zero $H$ background presented in section 3,
in section 4 we present our main claim regarding the relevance of Malcev algebras in string theory and then in sections 4 and 5 
we discuss the general structure of Malcev algebras in this context
as well as the relation between nonassociative and associative structures.
In particular, we generalize
nonassociative Malcev algebras that naturally appear  in
the quantum mechanics of systems with non-vanishing three-cocycles, such as a point particle moving in the field of a magnetic charge, to include electric as well as magnetic charges. These results find
their classical counterpart in the theory of Poisson-Malcev algebras \cite{poissonmalcev}, which we identify as the underlying nonassociative structures 
behind Stueckelberg's pioneering work on generalized Poisson brackets that do not obey the Jacobi identity.
In section 6 we also point our the relation between Malcev algebras and certain special geometries.
In the concluding section we stress that  
nonassociativity arises in the purely cubic form of Witten's
open string field theory \cite{cubic}
as pointed out by Horowitz and Strominger \cite{horstrom, strom, strom1},
as long as one
is not confined to the calculation of S-matrix elements
\cite{witten1, witten2, antal}.
The nonassociativity found in this context is relevant in
understanding the appearance of closed string states \cite{strom, strom1} in
a purely open string field theory.
In view of these facts, we close our paper with a brief discussion of the general role of nonassociativity in
string theory (see also the discussion in \cite{lust}), and in particular, we discuss the role of nonassociativity in closed string field theory \cite{strom, strom1},
which leads us to conjecture a string-field theoretic generalization of the AdS/CFT-like (holographic) duality.

\section{Stueckelberg's Generalization of Poisson brackets and Boltzmann's  H-theorem}
In this section we review Stueckelberg's generalization of Poisson brackets in classical statistical mechanics while preserving Liouville's  theorem and Boltzmann's H-theorem that follows from it \cite{ecgs}. Let $\xi^\alpha$ denote the coordinates in phase space with $\alpha, \beta , ...=1,2,..,2n$. Conservation of energy $H=H(\xi(t))$ implies that\begin{equation}
\dot{\xi}^\alpha(t)= \partial_t \xi^\alpha(t) = \Omega^{\beta \alpha}(\xi(t))\,  \partial_\beta H(\xi(t))
\end{equation}
where $\Omega^{\alpha\beta}(\xi) = -\Omega^{\beta\alpha}(\xi)$ is a symplectic "metric" in phase space. The scalar density $\mathfrak{\textbf{w}}$$ (\xi, t)$ of the Gibbs ensemble is positive semi-definite and satisfies the continuity equation\begin{eqnarray}
\int d^{2n}\xi \,\,  \mathfrak{\textbf{w}}(\xi,t) =1 \\
 \partial_t \, \mathfrak{\textbf{w}}(\xi,t) + \partial_\alpha ( \dot{\xi}^\alpha \mathfrak{\textbf{w}}(\xi,t) ) =0.
\end{eqnarray}
Under  coordinate transformations $ \xi^\alpha \rightarrow \xi^{'\alpha}$  in phase space $\mathfrak{\textbf{w}}$ transforms as
\begin{equation}
\mathfrak{\textbf{w}}\Longrightarrow \mathfrak{\textbf{w}}'(\xi',t) = |Det(\partial_\alpha \xi^{'\beta}(\xi))| \mathfrak{\textbf{w}}(\xi,t).
\end{equation}
Let $ dV(\xi)= g(\xi) d^{2n}\xi $ be the invariant volume element in phase space (with $g(\xi)$ representing the "density of volume"). Then
the invariant " scalar of the density" $\omega(\xi,t)$ is defined as
\begin{equation}
\omega(\xi,t) := \frac{\mathfrak{\textbf{w}}}{g} (\xi,t)\, > 0.
\end{equation}
Since $\Omega_{\alpha\beta}$ plays the role of metric in phase space Stueckelberg argues that the unique choice for $g$ is 
  \begin{equation} 
  g= | Det (\Omega^{\alpha\beta})^{-1/2}| \, >0. 
\end{equation} 
Then the continuity equation takes the form
\begin{eqnarray}
\partial_t \, \omega + D_\alpha ( \xi^\alpha \omega) = 0
 \end{eqnarray}
where $ D_\alpha = \partial_\alpha + \partial_\alpha (\log{g})$.
Now the  Liouville theorem states the "scalar  of density" must remain constant as the system evolves
\begin{equation}
\frac{d}{dt} \omega(\xi(t), t) := \dot{\omega}(\xi,t) = ( \partial_t \, \omega + \dot{\xi}^\alpha \partial_\alpha \omega )(\xi,t) =0 .
\end{equation}
This implies \begin{equation}
D_\alpha \dot{\xi}^\alpha =D_\alpha \left((\partial_\beta H) \Omega^{\beta\alpha} \right) = (\partial_\alpha \partial_\beta H ) \Omega^{\alpha\beta} + (\partial_\beta H ) D_\alpha \Omega^{\alpha\beta} =0
\end{equation}
which requires that the fundamental tensor $\Omega^{\alpha\beta}$ satisfy
\begin{equation}
D_\alpha \Omega^{\alpha\beta} = q^\beta =0.
\end{equation}
In terms of the density ${\mathcal{W}}^{\alpha\beta} = g \, \Omega^{\alpha\beta}$ it can be expressed as \begin{equation}
\partial_\alpha {\mathcal{W}}^{\alpha\beta}= q^\beta = 0
\end{equation} 
which has the form of the second set of Maxwell's equations with vanishing electric charges $q^\alpha$. 

Thus the time evolution of an observable $F(\xi)$ is given by\begin{equation}
\dot{F}(\xi) = \dot{\xi}^\alpha \partial_\alpha F(\xi)=\partial_\beta H \, \Omega^{\beta\alpha} \, \partial_\alpha F \equiv \{ H, F\} = -\{F , H \}
\end{equation}
where $\{F,H\}$ defines the generalized Poisson bracket. The Jacobian defined as 
\begin{equation}
J(F,G,H) \equiv \{F,\{G,H\}\} + \{H,\{F,G\}\}+ \{G,\{H,F\}\} 
\end{equation}
does not vanish in general under the generalized Poisson bracket. A simple calculation yields
\begin{equation}
J(F,G,H) = - 3 (\partial_\alpha F ) (\partial_\beta G) (\partial_\gamma H) \left(\Omega^{\rho[\alpha} \partial_\rho \Omega^{\beta\gamma]} \right).
\end{equation}
To have the Jacobi identity satisfied we must require
\begin{equation}
-3 \Omega^{\rho[\alpha } \partial_\rho \Omega^{\beta\gamma]} \equiv q^{\alpha\beta\gamma} =0
\end{equation}
which can be written in the form of the first set of Maxwell equations with vanishing magnetic charges
\begin{equation}
\partial_{[\alpha} \Omega_{\beta\gamma]} \equiv q_{\alpha\beta\gamma}=0 .
\end{equation} 
Only when the magnetic charges  $q_{\alpha\beta\gamma}$ vanish one can choose a polarization and bring the metric $\Omega^{\alpha\beta}$ to the canonical form locally:
\[
\begin{array}{|ccccc|}
0 & -1 & 0 & 0 & . \\
1 & 0 & 0 & 0 & . \\
0 & 0 & 0 & -1 & . \\
0 & 0 & 1 & 0 & .\\
. & . & . & . & . 
\end{array}
\]

Note that the above structure discovered by Stueckelberg fits beautifully into the Poisson-Malcev algebra discussed in
the mathematics literature \cite{poissonmalcev}.
An associative commutative algebra $A$ over a field  $F$ is called a Poisson-Malcev algebra if it is endowed with a $F$-bilinear map called
the Poisson-Malcev bracket $\{, \}$, which is antisymmetric $\{ f_1, f_2\} =- \{f_2, f_1\}$ and which satisfies the following conditions:
\begin{equation}
\{f_1 f_2, f_3\} = f_1\{f_2, f_3\} + \{f_1, f_3\}f_2
\end{equation}
i.e. the Leibnitz identity and 
\begin{equation}
J(f_1, f_2, \{f_1, f_3\}) = \{J(f_1, f_2, f_3), f_1\}
\end{equation}
for all $f_1, f_2, f_3 \in A$ where the Jacobian $J$ is defined as usual
\begin{equation}
J(f_1, f_2, f_3) \equiv \{\{f_1, f_2\}, f_3\} + \{\{f_2, f_3\}, f_1\} +\{\{f_3, f_1\}, f_2\} .
\end{equation}
We will see that this is precisely the classical limit of the quantum Malcev algebra to be discussed in section 4!

\subsection{Generalized Poisson brackets, minimal length and the stringy uncertainty principle}

The above analysis of Stueckelberg shows that Jacobi identity is not necessary to establish the H-theorem.
Furthermore Stueckelberg suggests that to the generalized Poisson brackets involving "magnetic charges" there must correspond a quantum theory in which the observables are no longer linear operators. This observation may have been one of the reasons why Stueckelberg considered the possible extension of quantum mechanics involving nonlinear operators. 
We should point out that nonassociativity corresponds to a particular kind of nonlinearity\footnote{In particular, nonassociativity precludes simple tensoring of free systems
and thus the physics of nonassociativity should be intrinsically interactive and therefore nonlinear.} and we  shall argue that Stueckelberg's attempts to find a nonlinear extension of quantum mechanics should be replaced by attempts to find a nonassociative extension of quantum mechanics. The minimum length that he introduced for a non-linear extension of quantum mechanics would then be related to nonassociativity. 

In particular Stueckelberg considered quantum theories with a "critical length" $\lambda_0 $  such that uncertainties in the measurements of coordinates satisfy
\begin{equation}
(\Delta X)^2 \eqslantgtr  ( \lambda_0 )^2 
\end{equation}
and proposed modifying the minimum uncertainty relation in one dimensional case as follows\footnote{In three dimensions he proposes  using the minimum uncertainty among all possible directions.}
\begin{equation}
(\Delta X)^2 (\Delta P)^2 = \frac{\hbar^2}{4} \left( 1 - \frac{(\lambda_0)^2}{(\Delta X)^2} \right)^{-1}
\end{equation}
which requires the modification of canonical commutation relations as
\begin{equation}
i [ P, X] = \hbar  \left( 1 - \frac{(\lambda_0)^2}{(\Delta
X)^2} \right)^{-1/2}.
\end{equation}

This relation is, at first, a bit different from
the canonical form of the string uncertainty principle \cite{stringyup, sup},
which usually reads as
\begin{equation}
\Delta X \Delta P \sim (1+ \alpha' \Delta P^2) \hbar.
\end{equation}
However, if we formally expand the inverse $ \left( 1 - \frac{(\lambda_0)^2}{(\Delta X)^2} \right)^{-1}$
and use to first order $\Delta X \sim \hbar \, \Delta P^{-1}$ we precisely get the stringy
uncertainty relation! (This formal procedure relates $\lambda_0$ with the string scale $ l_s$,
or equivalently with $\alpha' \sim l_s^2$.)
Both the physical and mathematical underpinnings of
the stringy uncertainty relation are still not clear, even though one can attempt
an algebraic formulation in terms of generalized $X$ and $P$ commutators (see the review in \cite{sup}). 
However, even at the classical level, at least naively, one is faced with certain interesting issues:
for example, one might say that by the Darboux theorem one can always redefine $X$ and $P$
so that the generalized Poisson bracket has the
canonical $\{X,P\} =1$ form. However, the Darboux theorem holds if the Jacobi identity is satisfied!
Another problem with this reasoning is that $X$ is constrained to be not smaller
than $\lambda_0$ (i.e. the string length scale $l_s$).
Thus the Poisson bracket between different $X$s has to be nontrivial
to reflect this fact.
(It is interesting that Stueckelberg does not comment about the $[X_i,X_j]$ commutator
because he is working in one dimension, which is trivial from the point of view of this commutator. In higher dimensions this commutator has to
be non-zero.) Then one has to close the structure and compute $[P_i,P_j]$
by imposing some suitable requirement. Usually one imposes the Jacobi identity (see the review in \cite{sup}) associated with the
Poisson structure, but
given Stueckelberg's reasoning one should impose the Poisson-Malcev structure and thus relate the magnetic
charge to the minimal length. This intuition makes sense in the string case because the $ H$ flux which is
responsible for nonassociativity   is related to the curvature of the metric by the string equations of motion \cite{lust}.
The connection of non commutativity and nonassociativity and the string uncertainty principle  was also investigated in \cite{lust}.
\section{Nonassociative star products and D-branes in curved backgrounds}

In order to introduce the concept of nonassociative star products in string theory
we follow the discussion originally presented in \cite{corn}. Note that the very important question
of the consistent string background was not addressed in \cite{corn}. This crucial question was cleared up in the more recent work \cite{lust}.
The basic set-up is that of open strings propagating in a curved background, following the
well-known construction of noncommutative field theories in open string theory \cite{connes, seibwitten, schom, chu,corn,ho}.
The sigma model action is ($2\pi \alpha' =1$)
\eq
S = {1 \over 2} \int g_{ab}(X) dX^a \wedge *dX^b
+ {i \over 2} \int B_{ab}(X) dX^a \wedge dX^b
\en
where the integration runs over the string worldsheet.
By turning on the $U(1)$ field at the boundary of the worldsheet one also has to include
\eq
S_b = \int F_{ab}dX^a \wedge dX^b.
\en
By going to the Riemann normal coordinates
and expanding the metric and the $B$ field around constant backgrounds
one gets, to leading order
\eq
S = {1 \over 2} g_{ab} \int dX^a \wedge *dX^b
+ i \int \omega
+{i \over 6} H_{abc} \int X^a  dX^b \wedge dX^c
\en
where the effective symplectic structure \cite{corn} 
\eq
\omega_{ab} (x) = B_{ab} + F_{ab} (x)
\en
and where the 3-form field strength is given by
\eq
H= dB.
\en
To leading order, i.e. in the weakly curved backgrounds, $H$ is constant.

The correlation function of operators inserted at the boundary in the case when $H=0$
are given by the well known result 
\eq 
<f_1 f_2...f_n> = \int V(\omega) dx (f_1 * f_2 *
...*f_n), 
\en 
where $V(\omega)$ is the appropriate volume form \cite{corn} and $*$ is the associative Weyl-Moyal-Kontsevitch star product with
respect to the symplectic structure
$\omega^{-1}$
\eq f * g = f g +{i \over 2}
\omega^{ab} \partial_{a}f \partial_{b} g+ ... 
\en 
The associativity is equivalent to the
fact that $\omega$ is closed, $d\omega =0$ \cite{corn}.

The original claim found in \cite{corn} is that if one considers the formal $g_{ab} \rightarrow 0$
limit above star product becomes nonassociative (note that the metric part $g_{ab}$  is actually
crucial for providing a consistent string background as shown in \cite{lust}). Thus formally one gets
\eq
f \bullet g = f g + {i \over 2}
{\tilde{\omega}}^{ab}
\partial_{a}f \partial_{b} g+ ...
\en
where, for constant $H$
\eq
{\tilde{\omega}}_{ab} (x) = B_{ab} + {1 \over 3} H_{abc} x^c + F_{ab} .
\en
Thus ${\tilde{\omega}}$ is not closed, $d {\tilde{\omega}} = H$,
and the $\bullet$ product is not associative
\eq
(f \bullet g) \bullet h - f \bullet (g \bullet h) =
{1 \over 6} {\tilde{\omega}}^{ia} {\tilde{\omega}}^{jb} {\tilde{\omega}}^{kc}
H_{abc} \partial_{i} f \partial_{j} g \partial_{k} h + ...
\en

From the world-volume point of view, this leads to non-commutativity
and nonassociativity of coordinates.
The commutator
\eq
[x^i, x^j]_{\bullet} \equiv x^i \bullet x^j -  x^j \bullet x^i
\en
is given by the following expression
\eq
[x^i, x^j]_{\bullet} = i {\tilde{\omega}}^{ij}.
\en
Furthermore, the Jacobi identity is violated
\eq
[x^i, [x^j, x^k]_{\bullet}]_{\bullet} +
[x^j, [x^k, x^i]_{\bullet}]_{\bullet} +
[x^k, [x^i, x^j]_{\bullet}]_{\bullet} = -
{\tilde{\omega}}^{ia}{\tilde{\omega}}^{jb}{\tilde{\omega}}^{kc} H_{abc}.
\en
This violation of the Jacobi identity is the starting point for our discussion of the
relevant nonassociative structures, Malcev algebras and their generalizations.
In the context of closed string theory in the constant $H$ background 
${\tilde{\omega}}_{ij} \sim H_{ijk} p^k$, $p$ being the momentum \cite{lust}, the resulting non-associative algebra is isomorphic to the Malcev algebra describing  the motion of
an electron moving  in the field of a constant magnetic charge distribution \cite{guzu}.  As pointed out in \cite{lust} the consistency of the conformal field theory involving the constant $H$ field 
crucially relies on the non-trivial gravitational background.  Hence we expect the more general "magnetic" charge distributions to be described by other Malcev algebras and their generalizations. 
Thus the nonassociative structure 
associated with the above violation of the Jacobi identity should be understood as a first step in
the construction of a more general nonassociative structure which is in some sense "generally covariantized"
in the presence of a non-trivial gravitational background \cite{chiadm}.

\section{Three-cocycles, nonassociativity and Malcev algebras}

In this section we point out that the nonassociative structure
encountered in the above violation of the Jacobi identity is well known in the
quantum mechanics of three-cocycles and is associated with the  mathematical
concept of Malcev algebras.

The canonical example is the quantum mechanics of a non-relativistic
electron moving in the background of a magnetic charge.
As pointed out long time ago by Lipkin, Weisberger and
Peshkin \cite{lipkin}, the commutators of the velocities of an
electron in the field of point-like magnetic monopole do not satisfy
the Jacobi identity at the position of the location of the 
magnetic monopole.

The basic commutation relations read
\eq
[q_a, q_b] =0, \quad  [q_a,v_b]=i\delta_{ab}
\en
and in particular
\eq
[v_a,v_b] =  i \epsilon_{abc} B_c
\en
where $a, b, c = 1,2,3$, and $\vec{B}$ denotes the magnetic field.
The commutators of the velocities yield the following
Jacobian-like quantity
\eq
[v_1, [v_2,v_3]] + [v_2, [v_3,v_1]] + [v_3, [v_1,v_2]] = - \vec{\nabla} \cdot \vec{B}
\en
and provide a classic example of
a non-vanishing three-cocycle as discussed in \cite{threecyc}.

Following  Jackiw's presentation in \cite{threecyc},
consider how the wave-function of the charge particle transform
under translations, a gauge-invariant action of which is
represented by the operator ($\hbar =1$)
\eq
U(\vec{a}) = \exp(i \vec{a} \cdot \vec{v}).
\en
The action of the translation group on the wave-function
$\Psi(\vec{q})$ of the electron
is then given by
\eq
U(\vec{a}) \Psi(\vec{q}) = \exp(i \vec{a} \cdot \vec{v})
\exp(-i \vec{a} \cdot \vec{p}) \Psi( \vec{q} + \vec{a}).
\en
Here the momentum $\vec{p}$
is $\vec{p} = \vec{v} + \vec{A}$ and $\nabla \times \vec{A} = \vec{B}$.
The three cocycle appears if one considers
\eq
[U(\vec{a_1})U(\vec{a_2})]U(\vec{a_3}) \Psi(\vec{q})=
\exp(i \alpha_3) U(\vec{a_1})[U(\vec{a_2}))U(\vec{a_3})] \Psi(\vec{q}).
\en
In the case of a magnetic monopole,
$\nabla \cdot \vec{B} = 4\pi \delta{(\vec{q})}$,
the three-cocycle is proportional
to the flux out of the tetrahedron formed from the three vectors
$\vec{a_1}, \vec{a_2}, \vec{a_2}$, with one vertex at $\vec{q}$.

Thus the three-cocycle arising in this situation is a signal
that the operators  representing  translations become nonassociative,
 so that the Jacobi identity fails.
(Thus, operators can no longer be linear operators acting in a Hilbert space.)

 It was pointed out  in \cite{bdz}, in any formulation of the
 quantum mechanical magnetic monopole problem in which the coordinates and the
 velocities of the electron are represented by operators acting on a Hilbert
 space, the Jacobi identity can not be violated, since such operators are
 always associative. The Jacobi identity can only be violated if the
 "representatives"  of coordinates and velocities are not operators in a Hilbert
 space, but belong to an intrinsically nonassociative algebra of observables.
 The  quantum mechanical description of a non-relativistic electron in the
 field of a magnetic charge distribution in terms of
 a nonassociative algebra of observables was studied by G\"{u}naydin and Zumino
 \cite{guzu}. They showed, in particular, that the nonassociative algebraic structure
 defined by velocities and coordinates of an electron in a {\it constant} magnetic
 charge distribution is  that of a Malcev algebra.

We now turn our attention to a brief review of the fundamental properties of
nonassociative algebras following \cite{schafer,jacob,guzu}.
The basic concept introduced in the study of nonassociative algebraic structures is that
of the associator. Given three elements $A, B, C$ of an algebra $\cal{A}$ their
associator $[A,B,C]$ is defined as 
\eq 
[A,B,C] \equiv (AB)C-A(BC). 
\en 
Obviously, the
associator vanishes for associative algebras. An alternative algebra is defined as an
algebra in which the following identities hold 
\eq 
[A,A,B] =0, \quad [B,A,A]=0. 
\en 
By
replacing $A$ by $A+C$ in these formulae one also gets 
\eq [A,C,B] + [C,A,B] = 0 \quad
[B,A,C] + [B,C,A] =0. \en 
Therefore, in an alternative algebra the associator $[A,B,C]$ is
an alternating function of its arguments $A,B,C$. The following, so-called Moufang
identities, can be  derived using the basic properties of the associator in an
alternative algebra  \cite{schafer,jacob,guzu}
\eq (ABA)C=A(B(AC)), \quad C(ABA)=((CA)B)A, \quad
(AB)(CA)=A(BC)A.
 \en
Note that  \eq (ABA)=(AB)A=A(BA) \en in alternative algebra.

Now, given an alternative algebra $\cal{A}$ one can define an algebra ${\cal{A}}^{-}$
with an  anti-symmetric product (i.e. commutator). The algebra ${\cal{A}}^{-}$ is not in
general a Lie algebra, because the Jacobi identity is not satisfied in ${\cal{A}}^{-}$.
The Jacobian is defined by
\eq J(A,B,C) = [A,[B,C]] + [B,[C,A]] + [C,[A,B]]
 \en
and is
proportional to the associator \eq J(A,B,C)= 6[A,B,C] \en which generally does not
vanish. The algebra ${\cal{A}}^{-}$ is a Malcev algebra \cite{malcev}. It is defined as an algebra with
an anti-symmetric product $\star$ \eq A \star B = - B\star A \en and a fourth order
identity, called Malcev identity \cite{malcev}
\eq 
(A \star B) \star (A \star C) = ((A \star B) \star
C) \star A + ((B \star C) \star A) \star A + ((C \star A) \star A) \star B . 
\en 
It can be
shown that the Malcev identity is equivalent to \eq J(A,B,A \star C) = J(A, B, C) \star A.
\en

The Malcev identity is trivially satisfied in the case of Lie algebras since the Jacobian
vanishes identically. Thus one can view Malcev algebras as a generalization of Lie
algebras. Malcev algebras arise naturally from alternative algebras under the commutator
product. One of the best known examples of an alternative algebra which is not
associative is the composition algebra of octonions. The seven imaginary units of the
octonions close under commutation and form a Malcev algebra which is the unique finite
dimensional simple Malcev algebra, up to isomorphisms, and which is not a Lie algebra.

Returning to the example of a non-relativistic electron moving in
the magnetic field
it can be now shown that the Malcev identity
\eq
J(v_1, v_2, [v_1,v_3]) = [J(v_1,v_2,v_3), v_1]
\en
implies that the divergence of the magnetic field, $\nabla \cdot \vec{B}$, must be $q$ independent.
Thus the algebra of velocities of a non-relativistic electron in the
constant magnetic field is a non-commutative and nonassociative
Malcev algebra. Since we have the usual canonical commutation relations between coordinates $q_i$ and $v_j$ the resulting Malcev algebra is infinite dimensional.

It is easy to see that the non-commutative and nonassociative algebra
of coordinates found in closed string theory in the presence of a constant $H$ background \cite{lust}, as discussed
at the end of section 3, is isomorphic to the algebra of velocities of a non-relativistic
electron in the field of a constant magnetic charge, and therefore, it is a Malcev algebra\footnote{Note that in the context of string theory $v$'s (velocities)  play the role of coordinates and $q$'s (coordinates) of the corresponding velocities (or momenta)!}.
However, as pointed out at the end of section 3, in the presence of a non-trivial
gravitational background that is needed for the consistent conformal field theory,
one needs to generalize the Malcev algebra by an appropriate ``general covariantization''.

One natural generalization of the above Malcev algebra can be obtained by putting velocities and coordinates on an equal footing by introducing "electric" charges that are dual to the already existing ``magnetic'' charges. The the full "phase space covariant" algebra with both ``electric'' and ``magnetic charges'' take the form:
\eqn
[q_a, q_b] = -i \epsilon_{abc} E_c , \quad  [q_a,v_b]=i\delta_{ab} \quad
{[}v_a,v_b {]} =  i \epsilon_{abc} B_c   \\
{[}v_1, {[}v_2,v_3{]}{]} + {[}v_2, {[}v_3,v_1{]}{]} + {[}v_3, {[}v_1,v_2{]}{]} 
= - \vec{\nabla} \cdot \vec{B}  \\
{[}q_1, {[}q_2,q_3{]}{]} + {[}q_2, {[}q_3,q_1{]}{]} + {[}q_3, {[}q_1,q_2{]}{]} 
= \vec{\nabla} \cdot \vec{E} .
\enn
The above  electric magnetic duality-covariant generalization of the algebra given in \cite{guzu} {\it does not} satisfy the Malcev identity and it represents a natural 
generalization of the Malcev algebra discussed in \cite{guzu}. Note that since this generalization requires both coordinates and velocities to be noncommuting we have to assume that both $E$ and $B$ fields 
must depend on coordinates as well as velocities and that is why the Malcev identity is not satisfied in general.

Finally, we stress that the Malcev algebra structure discussed in this section (following \cite{guzu}) without the "electric" sources  is precisely the quantum version of the classical Poisson-Malcev structure discussed in
section 2 in connection with Stueckelberg's classic paper \cite{ecgs}. The above generalization of the algebra of \cite{guzu} which includes both electric and magnetic charges corresponds to a covariant generalization of the Poisson-Malcev structure given in \cite{poissonmalcev}.
However our generalization given above requires that the symplectic metric in phase space given in equation (10) must not be covariantly constant. This in turn implies that the classical phase space formulation of Stueckelberg must be further generalized to phase spaces having non-vanishing "electric" as well as  "magnetic" sources. In this more general case one has to extend the implications of a minimal length to having both a minimal momentum as well as a minimal length. 

We should point out that the authors of \cite{lust} argued that the relevant structure for the central example discussed in this section of an electron   moving in the field of a magnetic charge  is the so-called twisted Poisson structure studied in 
\cite{twisted}.  It would be interesting to understand the connection between twisted Poisson structures and their quantization on one side and the Poisson-Malcev algebras and Malcev algebras on the other side.

In particular, given the
relevance of Malcev algebras (and their classical counterparts) in the context of more general
string backgrounds, it is natural to expect that  Malcev algebras (and their generalizations) find their natural realization in
string field theory. We will comment on this issue in the concluding
section of this paper.

\section{Malcev Algebras, 3-forms and exceptional Lie Algebras}

Since the operators acting on the Hilbert space of a quantum mechanical system are
associative, any intrinsically nonassociative algebra can not be realized by such
operators. This would of course not be true for any quantum mechanical system that has no
Hilbert space formulation. As pointed out in the introduction there is a unique quantum mechanical system  that
has no formulation over an Hilbert space. This is the octonionic quantum mechanics which
was shown to satisfy all the axioms of quantum mechanics within the Jordan density matrix
formalism of Jordan and which has no Hilbert space realization \cite{gpr}. Later the Jordan
formulation of quantum mechanics was generalized to the {\it quadratic} Jordan
formulation which extends to the octonionic quantum mechanics as well as allowing one to
define quantum mechanics over finite fields \cite{mg}. 

The algebraic structure
underlying the octonionic quantum mechanics is the exceptional Jordan algebra
$J_3^{\mathbf O}$ defined by $3\times 3 $ hermitian octonionic matrices. The Jordan
algebras are defined by a symmetric product
\eq
A \cdot B = B \cdot A
\en
and the Jordan identity
\eq
A \cdot( B \cdot  A^2) = ( A \cdot B ) \cdot A^2 .
\en
The algebra $J_3^{\mathbf O}$ is the unique Jordan algebra that has no realization in
terms of associative matrices with the Jordan product being one half the anticommutator.
Even though the algebra  $J_3^{\mathbf O}$ has no realization in terms
of associative matrices it still admits an embedding into a Lie algebra
that has realization in terms of associative matrices. In fact all
Jordan algebras admit such an embedding into a Lie algebra. Conversely
one can construct  Lie algebras from Jordan algebras through
what is known as Tits-Koecher-Kantor (TKK) construction \cite{tkk}.

Consider a 3-graded Lie algebra $g$:
\begin{equation}
g = g^{-1} \oplus g^{0} \oplus g^{+1}
\end{equation}
where $ \oplus $ denotes vector space direct sum and $g^{0}$ is a subalgebra of maximal
rank. We have the formal commutation relations of the
 elements of various grade subspaces
\begin{equation}
[g^{m},g^{n}] \subseteq g^{m+n} \; ; m,n=-1,0,1
\end{equation}
where $ g^{m+n} = 0 $ if $|m+n|>1$. Every simple Lie algebra with such a 3-graded
structure can be constructed in terms of an underlying Jordan triple system (JTS) $V$ via
the TKK
construction \cite{tkk}. This construction establishes a one-to-one mapping
between the grade $+1$ subspace of $g$ and the underlying JTS $V$:
\begin{equation}
U_{a} \in g  \Longleftrightarrow  a \in V .
\end{equation}
Every such Lie algebra $g$ admits
 a conjugation (involutive automorphism) $\dagger$
under which the elements of the grade $+1$ subspace get mapped into the elements of the
grade $-1$ subspace.
\begin{equation}
U^{a} = U_{a}^{\dagger}  \in g^{-1}.
\end{equation}
One then defines
\begin{equation}
\begin{array}{l}
[U_{a},U^{b}] = S_{a}^{b} \\
{[}S_{a}^{b}, U_{c}{]} = U_{(abc)}
\end{array}
\end{equation}
where $S_{a}^{b} \in g^{0}$ and $(abc)$ is a triple product under which the
 elements of $V$ close. Under conjugation $\dagger$ one finds
\begin{equation}
\begin{array}{l}
(S_{a}^{b})^{\dagger} = S_{b}^{a} \\
{[}S_{a}^{b},U^{c}{]} = - U^{(bac)} .
\end{array}
\end{equation}
The Jacobi identities in $g$ are satisfied if and only if the ternary product $(abc)$
satisfies the defining identities of a JTS:
\begin{equation}
\begin{array}{l}
(abc) = (cba)  \\
(ab(cdx))-(cd(abx))-(a(dcb)x)+((cda)bx) = 0 .
\end{array}
\end{equation}
The generators $S_{a}^{b}$ belonging to the grade zero subspace form a
 subalgebra
which is called the structure algebra of $V$ :
\begin{equation}
[S_{a}^{b},S_{c}^{d}] = S_{(abc)}^{d}-S_{c}^{(bad)}=S_{a}^{(dcb)} -S_{(cda)}^{b}.
\end{equation}

The exceptional Lie algebras $G_{2}, F_{4}$ and $E_{8}$ do not admit a TKK type
construction. A generalization of the TKK construction to more general triple systems was
given by Kantor \cite{IK}. All finite dimensional simple Lie algebras admit a
construction over these generalized triple systems which we call Kantor triple systems
(KTS). Kantor's  construction  of Lie algebras was generalized to a unified
construction of Lie and Lie superalgebras in \cite{BG79}.

Kantor construction starts from the fact that every simple Lie algebra $g$ admits  a
5-grading (Kantor structure) with respect to some subalgebra $g^{0}$ of maximal rank
\cite{IK,BG79}:
\begin{equation}
g = g^{-2} \oplus g^{-1} \oplus g^{0} \oplus g^{+1} \oplus g^{+2}.
\end{equation}
One associates with the grade $+1$ subspace of $g$ a triple system $V$ and labels the
elements  of $g^{+1}$ subspace with the elements of $V$ \cite{IK,BG79}:
\begin{equation}
U_{a} \in g^{+1} \Longleftrightarrow a \in V.
\end{equation}
Every simple Lie algebra $g$ admits a conjugation  which maps the grade $+m$ subspace
into the grade $-m$ subspace. Therefore one can also label the elements of the grade $-1$
subspace by the elements of $V$ :
\begin{equation}
U^{a} \equiv U_{a}^{\dagger} \in g^{-1} \Longleftrightarrow U_{a} \in g^{+1}.
\end{equation}
One defines the commutators of $U_{a}$ and $U^{b}$ as
\begin{equation}
\begin{array}{l}
{[}U_{a} , U^{b}{]} = S_{a}^{b} \in g^{0}  \\
{[}U_{a} , U_{b}{]} = K_{ab} \in g^{+2}               \\
{[}U^{a} , U^{b}{]} = K^{ab} \in g^{-2}             \\
{[}S_{a}^{b} , U_{c}{]} = U_{(abc)} \in g^{+1}
\end{array}
\end{equation}
where (abc) is the triple  product under which the elements of $V$ close. The remaining
non-vanishing commutators of $g$ can all be expressed in terms of the triple product (abc):
\begin{equation}
\begin{array}{l}
{[}S_{a}^{b} , U^{c}{]} = - U^{(bac)}           \\
{[}K_{ab} , U^{c}{]} = U_{(acb)} - U_{(bca)}  \\
{[}K^{ab} , U_{c}{]} = - U^{(bca)} + U^{(acb)}\\
{[}S_{a}^{b} , S_{c}^{d}{]} = S_{(abc)}^{d} - S_{c}^{(bad)}      \\
{[}S_{a}^{b} , K_{cd}{]} = K_{(abc)d} + K_{c(abd)} \\
{[}S_{a}^{b} , K^{cd}{]} = - K^{(bac)d} - K^{c(bad)} \\
{[}K_{ab} , K^{cd}{]} = S_{(acb)}^{d} - S_{(bca)}^{d} - S_{(adb)}^{c} + S_{(bda)}^{c}.
\end{array}
\end{equation}
The Jacobi identities of $g$ follow from
 the following identities \cite{IK,BG79}
\begin{equation}
\begin{array}{l}
(ab(cdx)) - (cd(abx)) - (a(dcb)x) + ((cda)bx) = 0 \\
\{ (ax(cbd)) - ((cbd)xa) + (ab(cxd)) + (c(bax)d) \} - \{ c \leftrightarrow d \} = 0
\end{array}
\end{equation}
which we take as the defining identities of a KTS. In general a given simple Lie algebra
can be constructed in several different ways by the above method corresponding to
different choices of the subalgebra $g^{0}$ and different ternary algebras.

In many instances ternary algebras can be defined in
terms of an underlying "binary" algebras. For example one can define a triple product
over a Jordan algebra that satisfies the two conditions defining a Jordan triple system
as follows:
\eq
(abc)\equiv a \circ (b \circ c) + c \circ (b \circ a) - ( a \circ c) \circ b = (cba) .
\en
In those instances when the ternary algebras are defined by ordinary algebras, one can reverse the process of TKK or
Kantor construction to define algebras from Lie algebras.\footnote{ Similarly,
one can give a vertex operator construction of underlying algebras and triple
systems and their affine extensions starting from affine Lie algebras and
super-algebras\cite{affinevertex}.}    In fact this is how Jordan
superalgebras were defined and classified by Kac \cite{kac}.
Below we give more examples of such embeddings  and discuss the case of a simple  Malcev algebra of dimension seven defined by imaginary octonions. 

The division algebras $ \mathbf{A} = \mathbf{R}, \mathbf{C}, \mathbf{H}$ and $
\mathbf{O}$ and their tensor products with each other define KTS's under the ternary
product
\eq
(abc) =a\cdot (\overline{b}\cdot c) + c\cdot (\overline{b}\cdot a) - b\cdot
(\overline{a}\cdot c)
\en
where $a,b,c$ are elements of $ \mathbf{A} \times \mathbf{A'} $ and the bar denotes
conjugation in the underlying division algebras. Using the Kantor construction above one
obtains the Lie algebras of the famous Magic Square. In Table 1. we list these algebras
and the corresponding Lie algebras.
\begin{table}[ht]
 \caption{~}
\begin{center}
\begin{displaymath}
\begin{array}{|c|c|c|c|c|}
\hline
\mathbf{A}\times \mathbf{A'}& \mathbf{R} & \mathbf{C} &\mathbf{H}&\mathbf{O} \\
\hline
\mathbf{R} & SO(3) & SU(3) & USp(6)& F_4 \\
\mathbf{C} & SU(3) & SU(3)^2 & SU(6) & E_6 \\
\mathbf{H}& USp(6) & SU(6) & SO(12) & E_7 \\
\mathbf{O} & F_4 & E_6 & E_7 & E_8 \\
\hline
\end{array}
\end{displaymath}

\end{center}
\end{table}

Consider the construction of the exceptional Lie algebra $F_4$ over the division algebra $\mathbb{O}$ of octonions.
The triple product in that case is 
$(abc) =a\cdot (\overline{b}\cdot c) + c\cdot (\overline{b}\cdot a) - b\cdot
(\overline{a}\cdot c) $. Let $e_0$ be the identity element of $\mathbb{O}$ and $e_A$ be the seven imaginary units.
Consider the elements $U_{e_A}$ of the Lie algebra $F_4$ belonging to grade $+1$ space labelled by $e_A$. Then define the product $\star$ between two elements of grade $+1$  space via the double commutator 
\eq
U_{e_A} \star U_{e_B} \equiv  [[U_{e_A} , U_{e_B}],U^{e_0}] = U_{[e_A,e_B]}. 
\en
Then under this product the grade $+1$ elements $U_{e_A}$ of $F_4$ form a Malcev algebra of dimension seven.

\subsection{Exceptional Lie algebras and anti-symmetric tensors of rank three}

Since the three form field strengths play a fundamental role in the results presented in this paper as well as in \cite{lust} let us investigate their algebraic properties. Clearly one can not define a binary product over the space of tensors of rank three.
However one can define a triple product among them such that they close under it. In fact,
exceptional Lie algebras of the $E$-series can be constructed in a unified manner over
triple systems defined by antisymmetric tensors of rank three in various dimensions
\cite{IK}. Let $x_{ijk}, y_{ijk}, z_{ijk} , (i,j,...=1,2,...,n )$ be totally
antisymmetric tensors of rank three in $n$ dimensions. Define a ternary product among
such tensors as
\eq
(xyz)_{ijk} = y_{imn} x_{pmn} z_{pjk} + y_{pjn} x_{pmn} z_{imk} + y_{pmk}x_{pmn} z_{ijn}
- \frac{1}{3} y_{mnp} x_{mnp} z_{ijk}.
\en
For $n=6,7$ this triple product satisfies the defining conditions of a KTS and lead to
the construction of $E_6$ and $E_7$, respectively.
\eq
\begin{array}{cccccc}
E_{6}=& T^-  \oplus & T_{ijk}  \oplus & T^i_j \oplus & T^{ijk}  \oplus & T^+ \\
78 = & \bar{1} & \bar{20} & U(6) & 20 & 1
\end{array}
\en
\eq
\begin{array}{cccccc}
E_{7}=& T_i  \oplus & T_{ijk}  \oplus & T^i_j \oplus & T^{ijk}  \oplus & T^i\\
133 = & \bar{7} & \bar{35} & U(7) & 35 & 7
\end{array}
\en
For $n=5$ and $n=4$ the above triple product satisfies the defining identities of a JTS
and lead to the construction of $E_5=SO(10)$ and $E_4=SU(5)$, respectively.
\eq
\begin{array}{cccc}
E_{5}= & T_{ijk}  \oplus & T^i_j \oplus & T^{ijk} \\
45 = & \bar{10} & U(5) & 10
\end{array}
\en
\eq
\begin{array}{cccc}
E_{4}= & T_{ijk}  \oplus & T^i_j \oplus & T^{ijk} \\
24 = & \bar{10} & U(5) & 10
\end{array}
\en
For $n=8$ the generalized Kantor construction leads to the construction of the largest
exceptional Lie algebra $E_8$ with a seven graded structure.
\eq
\begin{array}{cccccccc}
E_{8}=& T_i  \oplus&T^{ij} & T_{ijk}  \oplus & T^i_j \oplus & T^{ijk}  \oplus &T_{ij} & T^i\\
248 = & \bar{8} & 28 & \bar{56} & U(8) & 56 & \bar{28} & 8
\end{array}
\en
In nine dimension the generalized Kantor construction  leads to the
construction of the affine Lie algebra $E_9$. It is tantalizing to speculate that a
further generalization of the Kantor construction leads to the construction of the
hyperbolic Lie algebras $E_{10}$ and $E_{11}$  in terms of antisymmetric tensors in ten and eleven dimensions, respectively.

One can reverse the above construction of exceptional Lie algebras and define the triple systems  of antisymmetric tensors of rank three in dimensions $d<9$ corresponding to  a subspace of $E_d$  involving double commutators as was done for the simple Malcev algebra of dimension seven. 

\section{Simple Malcev algebra $\mathbb{O}^-$ and related geometric structures}

In this section we point out that the nonassociative algebraic structures discussed above
appear naturally in the context of certain special geometries.
The relation between 
 between the  6-manifolds with  the
$G_2$ invariant almost complex structures and the simple Malcev algebra $\mathbb{O}^-$  was investigated in the mathematics literature long time ago \cite{calabi,frolicher,j1,j2, exg}. 
These geometries can be naturally associated with new backgrounds of string theory in which a non-constant $H$ field is turned on.

In particular, as discussed in detail in \cite{muratnick}, the almost complex structure and the associated
torsion tensor on $S^6$ are intimately related to octonions. Here we review the basic facts about it closely following \cite{muratnick}.
An almost complex structure of an even dimensional manifold $M$ is a mixed tensor $F_k^i$ which satisfies
\eq
F_k^i F_j^k = - \delta_j^i
\en
The torsion of an almost complex structure is defined as \cite{frolicher}
\eq
\tau_{jk}^i \equiv \frac{1}{2} (A_{mj}^i F_k^m - A_{mk}^i F_j^m)
\en
where $A_{jk}^i$ is given by the following expression
\eq
A_{jk}^i \equiv \frac{1}{2}(\nabla_j F_k^i - \nabla_k F_j^i)
\en
where $\nabla_i$ is a covariant derivative on $M$ with respect to some symmetric affine
connection. The symmetry of the affine connection implies that
\eq
A_{jk}^i \equiv \frac{1}{2}(\partial_j F_k^i - \partial_k F_j^i)
\en
Note that the almost complex structure $F_i^j$ is integrable to a complex structure provided
the torsion tensor $\tau_{ij}^k$ is vanishes \cite{frolicher, calabi}.

As emphasized in \cite{muratnick} an almost complex structure of $S^6$ can be defined using the imaginary Cayley numbers \cite{frolicher, calabi, j1},
by thinking of $S^6$ as a hypersurface in the space of imaginary octonions, also known as Cayley space $I^7$ \cite{j2}.
A vector $X$ in $I^7$ can be represented as ($A=1,..,7$)
\eq
X = X^A e_A
\en
where $e_A$ are the imaginary units of octonions \cite{muratfeza}.
The natural scalar product of two vectors $X$ and $Y$ in $I^7$ reads as follows
\eq
(X,Y) \equiv \frac{1}{2}(\bar{X} Y + X\bar{Y}) = X^A Y_A
\en
where bar denotes octonion conjugation $e_A \to - e_A$.
The natural cross product of two vectors is given by the commutator
\eq
X \wedge Y \equiv \frac{1}{2}[X Y - X Y]= - Y \wedge X.
\en
Under this product the space $I^7$ becomes the unique simple finite dimensional Malcev algebra defined by the imaginary octonions. 

Let us, following \cite{muratnick}, denote the unit normal to $S^6$ as $\hat{n}$ and choose a set of basis vectors $\hat{e}_i$ in the tanget
space so that the metric $g_{ij}$ on $S^6$ is
\eq
g_{ij} = (\hat{e}_i, \hat{e}_j) .
\en
Then the almost complex structure on $S^6$ can be defined via
\eq
\hat{e}_i \wedge \hat{n} = F_i^j \hat{e}_j .
\en
From the properties of  the cross product one can verify that indeed $F_k^i F_j^k = - \delta_j^i$ \cite{muratnick}.
Using the cross product one can  also define a mixed tensor $T_{ij}^k$ 
\eq
\hat{e}_i \wedge \hat{e}_j = - F_{ij} \hat{n} + T_{ij}^k \hat{e}_k .
\en
Note that $T_{ijk} = T_{ij}^m g_{mk}$ is completely antisymmetric in its indices
\eq
T_{ijk} = (\hat{e}_i \wedge \hat{e}_j, \hat{e}_k) = - T_{jik} = T_{kij}
\en
and that $T_{ijk} T^{ljk} = 4 \delta_i^l$.
By utilizing the equations of Gauss and Weingarten for the covariant derivatives $\nabla_i$
\eq
\nabla_i \hat{e}_j = H_{ij} \hat{n}, \quad \nabla_i \hat{n} = -H_i^j \hat{e}_j ,
\en
where $H_{ij}$ is the second fundamental tensor, 
and the Gauss-Codazzi equations for the curvature $R_{ijkl} = H_{il} H_{jk} - H_{jl} H_{ik}$ and
$\nabla_k H_{ji} = \nabla_j H_{ki}$, one can show that for $S^6$ \cite{muratnick}
\eq
H_{ij} = \lambda g_{ij}.
\en
By covariant differentiation of  $\hat{e}_i \wedge \hat{n} = F_i^j \hat{e}_j$ one finds
\eq
\nabla_j F^k_i = H_j^m T_{mi}^k. 
\en
Furthermore by covariantly differentiating $\hat{e}_i \wedge \hat{e}_j = - F_{ij} \hat{n} + T_{ij}^k \hat{e}_k$ one obtains on $S^6$
\eq
\nabla_k T_{ij}^k= - 4\lambda F_{ij}
\en
as well as
\eq
\nabla_k F_{ij} = T_{ijk}.
\en
The almost complex structure
$F_{ij}$
 is covariant under 
the automorphism group $G_2$ of octonions and 
the relevant nonassociative algebraic structure is  the Malcev algebra $\mathbb{O}^-$ defined by the  seven imaginary units of
octonions.

In the context of our note, the almost complex structure $F_{ij}$ can obviously play the role of the $B$ field and
the torsion $T_{ijk}$ of the associated non-vanishing $H$ field.  As such they correspond to a  background with non-constant $H$ and $B$ fields defined by an underlying Malcev algebra \footnote{ The coset space $G_2/SU(3)$ is one of the spaces considered  in \cite{Lust:1986ix} on compactifications of superstrings on six dimensional  Ricci flat coset spaces and, more recently, in the work of \cite{lustads} on backgrounds  that lead  to four dimensional anti-de Sitter space times.}.

\section{Discussion: A generalized AdS/CFT-like duality}

In this paper we first reviewed the work of Stuckelberg  on the generalization of Poisson brackets in classical mechanics while preserving Boltzmann's H-theorem that , in general, violate the Jacobi identity. We proposed that the nonlinear extension of the corresponding quantum mechanics envisaged by Stueckelberg, which involves the fundamental length, should be formulated as a nonassociative extension of quantum mechanics. We found that the relevant classical structure is the Poisson-Malcev algebra and  its generalizations.   

In the context of string theory the nonassociative
algebraic structures  appear in the description of
closed strings in curved backgrounds \cite{lust}. We  pointed out
that under certain assumptions these nonassociative structures coincide
with nonassociative Malcev algebras which naturally appear in
the quantum mechanics of systems with non-vanishing three-cocycles, such as an electic point particle moving in the field of a magnetic charge.
We also generalized the Malcev algebra of a point particle moving in the field of a magnetic charge to include both  electric and magnetic charges.

In  recent work \cite{lust} the role of nonassociativity was explored in 
closed string theory involving non-trivial gravitational backgrounds.  One of the hallmarks of quantum gravity is holography. Given the holographic
formulation of string theory in asymptotically AdS backgrounds, it is natural to ask whether a generalized
AdS/CFT-like dictionary \cite{adscft} can be formulated involving
nonassociative structures in closed string theory, by relating them to noncommutative yet associative
structures in open string theory. (As pointed out in section 4, the nonassociative structures 
can in general be 
embedded into Lie algebras. The question is whether this purely
algebraic relation between nonassociative and associative algebraic structures can be extended to include dynamics.)

To address this possibility,  we first note that
the failure of the Jacobi identity of the form discussed in this
paper is also known to occur in purely cubic open string field
theory \cite{horstrom}. This associativity anomaly can
be demonstrated by an oscillator calculation in purely cubic
string theory. The technical reason for the appearance of this
associativity anomaly can be understood as follows:
one can think of operators appearing in open string field theory
in terms of infinite dimensional matrices. For a class of
infinite dimensional matrices single sums appearing in the
usual product of two matrices are absolutely convergent, yet
double sums, appearing in the product of three matrices are not absolutely convergent,
leading to associativity anomalies.

It is claimed in the literature that this nonassociativity
in purely cubic open string field theory
is responsible for the appearance of closed string states \cite{horstrom}.
In particular, Strominger \cite{strom, strom1},
has proposed a gauge invariant cubic action
describing bosonic closed string field theory, in which the
basic dynamical objects are open strings. The action is given by the
associator for the string field product, which due to the
associativity anomaly is non-vanishing.

Very roughly, Strominger's closed string field theory action looks like its open string field
theory counterpart, i.e. it is cubic
\eq
S_c{(\Psi)} = \int \Psi \times (\Psi \times \Psi)
\en
except for the nonassociative nature of the $\times$ product. The $\Psi$ field is also of
the ``open string'' type.
The open string field theory of Witten \cite{witten1, witten2}
can be also written in a purely cubic form  (see \cite{cubic, antal})
\eq
S_o{(\Phi)} = \int \Phi * \Phi * \Phi
\en
where $*$ is a non-commutative but associative product and where $\Phi$ is
an open-string field.
Given the fact that the Laplace transform of an exponential of a cubic
(i.e. the Airy function)
in the WKB limit becomes an exponent of a square root of a cubic,
$ \int dx \exp{(t x - x^3)} \sim \exp{(-t^{3/2})}$,
(where $t$ obviously scales as $x^2$)
one is tempted to conjecture the following AdS/CFT-like ``holographic"
dictionary between the generating functional of the cubic open string field theory at large coupling 
and its cubic closed string field theory counterpart
\eq
\langle \exp{(\int J \Phi)} \rangle_{o} \equiv Z[J] \to \exp{[-S_c(\Psi)]}.
\en
Here $\Psi \to J$ in the region of the closed string Hilbert space on which the
3-cocycle anomaly vanishes. (Note that $\Psi \sim J$ scales as $\Phi^2$).
Thus the on-shell action for Strominger's nonassociative cubic closed string field theory would
compute the generating functional for Witten's noncommutative cubic open string field theory at large coupling.
In this sense this would be a natural non-geometric generalization of the AdS/CFT-like dictionary\footnote{This conjecture is , in some sense,  made more plausible by the existence of the Legendre invariant cubic forms \cite{GST,boris,ltcubic,kazpol}. Remarkably, the C-tensors of maximal supergravity as well as the exceptional supergravity define cubic forms that are Legendre invariant \cite{ltcubic}. }.

Given the robust nature of Malcev algebras (and their generalizations), as discussed in sections 4 and 5 of this paper, and given the fact that associative anomalies do appear in the context of
string field theory, it is natural to conjecture that the relevant algebraic structure represented by the non-associative $\times$ product in Strominger's cubic
formulation of closed string theory is of a Malcev algebra type\footnote{The closed string fields might be realized in the context of the quadratic Jordan formulation.}. Its explicit realization is an open and fascinating fundamental question.
Nevertheless, we hope that the discussion presented in this article makes a compelling case for the fundamental role of nonassociativity in
string theory.

{\bf Acknowledgements:}
This paper had a 12 year long gestation. We benefited from discussions with R. Gopakumar, A. Strominger and B. Pioline during the initial period. More recently we benefited from discussions with L. Freidel, R.G. Leigh, M. Krusczenski and Sung-Sik Lee and especially, D. L\"ust.
We also wish to acknowledge the kind hospitality of Harvard University, Caltech, USC, the Institute for Advanced Study at Princeton and the Miami Winter conference where parts of this work were initiated and developed.  The work of M.~G.\ was supported in part by the National Science Foundation under
Grant Number PHY-1213183 and PHY-0855356.  The work of D.~M.\ was supported in part by the US Department
of Energy under grant number DE-FG03-84ER40168.

\end{document}